\documentclass[aps,prl,twocolumn]{revtex4}
\usepackage{graphics,graphicx,amssymb,amsmath,epstopdf}
\usepackage{float}

\begin{document}

\title{Inelastic x-ray scattering measurements of phonon dynamics in URu$_2$Si$_2$}
\author{D.~R.~Gardner$^{1}$}
\author{C.~J.~Bonnoit$^{1}$}
\author{R.~Chisnell$^{1}$}
\author{A.~H.~Said$^{2}$}
\author{B.~M.~Leu$^{2}$}
\author{T.~J.~ Williams$^{3,4}$}
\author{G.~M.~Luke$^{3}$}
\author{Y. S.~Lee$^{1}$}
\affiliation{$^{1}$Department of Physics, Massachusetts Institute of Technology, Cambridge, MA 02139}
\affiliation{$^{2}$Advanced Photon Source, Argonne National Laboratory, Argonne, Illinois, USA}
\affiliation{$^{3}$Department of Physics and Astronomy, McMaster University, 1280 Main Street West, Hamilton, ON, Canada L8S 4M1}
\affiliation{$^{4}$Quantum Condensed Matter Division, Neutron Sciences Directorate, Oak Ridge National Lab, Oak Ridge, TN, 37831, USA}
\date{\today}

\begin{abstract}
We report high-resolution inelastic x ray scattering measurements of the acoustic phonons of URu$_2$Si$_2$. At all temperatures, the longitudinal acoustic phonon linewidths are anomalously broad at small wave vectors revealing a previously unknown anharmonicity. The phonon modes do not change significantly upon cooling into the Hidden Order phase. Our data suggest that the increase in thermal conductivity in the Hidden Order phase cannot be driven by a change in phonon dispersions or lifetimes. Hence, the phonon contribution to the thermal conductivity is likely much less significant compared to  that of the magnetic excitations in the low temperature phase.\end{abstract}

\maketitle

%Since the initial work almost three decades ago, considerable theoretical and experimental work has still failed to reveal the nature of the Hidden Order (HO) transition in the heavy fermion URu$_2$Si$_2$\cite{Palstra:1985uc,Maple:1986uy}. A sharp specific heat peak at $T_o=17.5K$ evidences a phase transition that's order parameter remains unknown\cite{Mydosh:2011go,Mydosh:2014fk}.

A full thirty years has passed since the Hidden Order phase in the heavy fermion URu$_2$Si$_2$ that occurs below $T_0=17.5K$ was first discovered.\cite{Palstra:1985uc,Maple:1986uy} For all the concerted effort, there is still no clear understanding of the ordering of this phase.\cite{Mydosh:2011go,Mydosh:2014dq,Chandra:2013gv,Pepin:2011jc,Varma:2006ea,Mineev:2005hl,Dubi:2011cs,Kung:2015ip,Ikeda:2012jn,Suzuki:2014fi,Elgazzar:2009cx} The Hidden Order (HO) phase is characterized by a gapping out of parts of the Fermi surface, a large Ising anisotropy with the easy magnetic axis c in the tetragonal structure, and a gapping out of c-axis polarized magnetic fluctuation.

Above $T_0$, URu$_2$Si$_2$ is body-centered tetragonal (space group $I4/mmm$) . Below $T_0$, there is some uncertainty; some work suggest that electronic structure has the same symmetry as the antiferromagnetic state, which occurs under pressure, and reduces the symmetry to simple tetragonal\cite{Hassinger:2010wk}. Torque magnetometry also suggests a breaking of C4 rotational symmetry in the basal-plane magnetic susceptibility, further reducing the symmetry\cite{Okazaki:2011ec}. Cyclotron resonance \cite{Tonegawa:2013vv, Tonegawa:2012ct} and NMR data \cite{Kambe:2013el} are supportive of C4 symmetry breaking, though the interpretation remains controversial. For this analysis, we will use the body-centered tetragonal structure. %Figure \ref{Figure:Structure} shows corresponding Brillouin zone with high symmetry points labeled.

\begin{figure}[!ht]
\includegraphics[width=8cm]{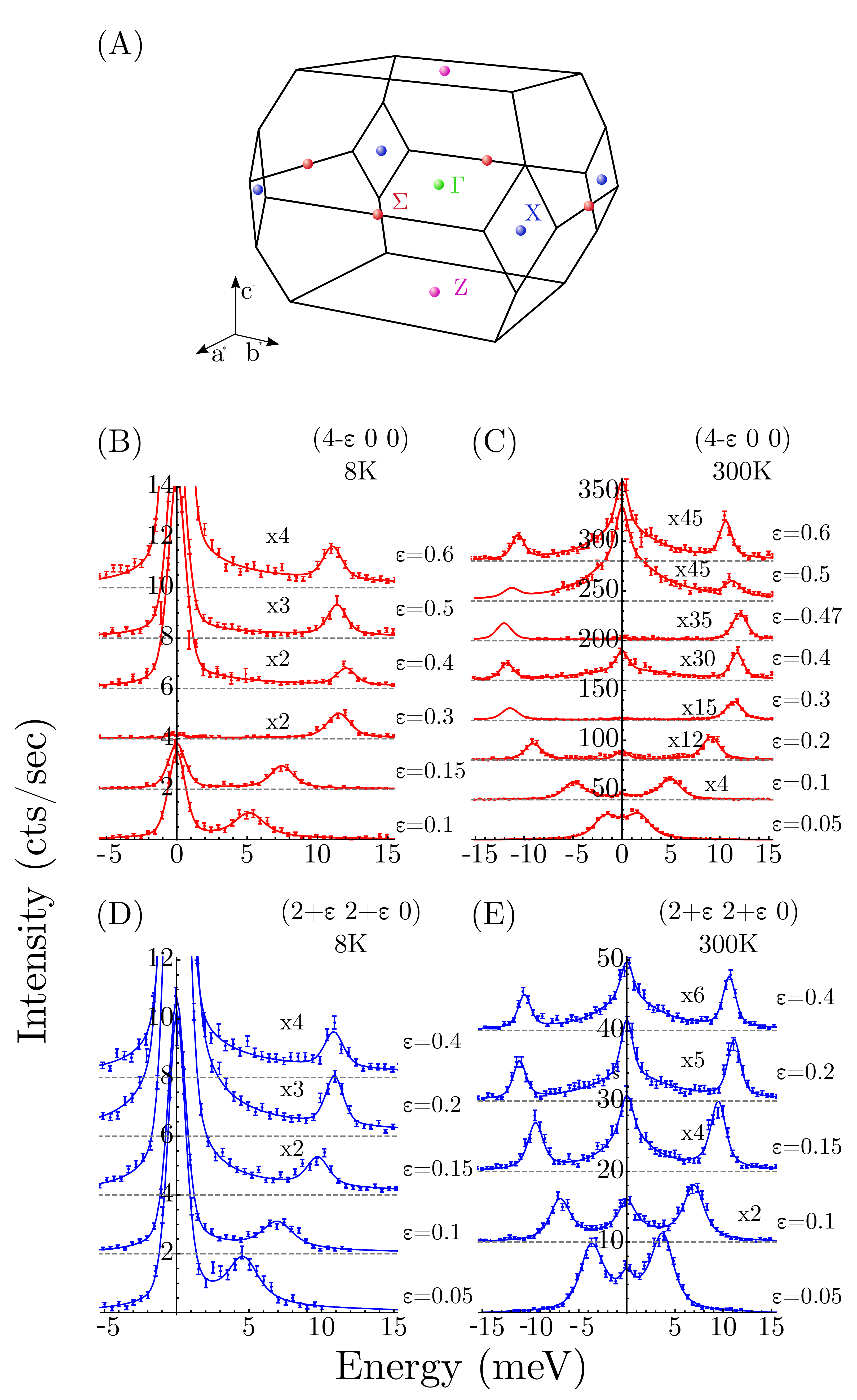}
\caption{(A) The Brillouin zone for the body centered tetragonal cell. (B-E) Phonon measurements with fits as described in the text. (B) and (C) show longitudinally polarized acoustic phonon modes along the $\Gamma$ to $\Sigma$ at  8K and 300K. (D) and (E) show longitudinal acoustic modes along $\Gamma$ to $X$ at 8K and 300K. For clarity, successive spectra where shifted by 2, 50, 2, and 10 cts/sec respectively.}
\label{Figure:Scans}
\end{figure}

Previous experimental work suggests the HO transition couples to the lattice. At $T_0$, the c-axis thermal expansion coefficient is negative while the in-plane coefficient peaks sharply\cite{deVisser:1986wz}. Thermal transport data show an increase in thermal conductivity that is thought to be dominated by a change in phonon thermal conduction\cite{Behnia:2005iy,Sharma:2006fa}. Ultrasound measurements show a softening of the ${(C_{11}-C_{12})/2}$ elastic constant, which is associated with an orthorhombic strain field\cite{Yanagisawa:2012gg}. Finally, some x-ray diffraction data show an orthorhombic distortion only in crystal samples that have high residual resistivity ratio (RRR.) \cite{Tonegawa:1jm}\cite{Tabata:2014is}.

%\begin{figure}
%\includegraphics[width=8.5cm]{Figures/Structure/structure_v4.png}
%\caption{Structure of URu$_2$Si$_2$. (A) The body centered tetragonal unit cell. (B) $\theta$-$2\theta$ diffraction data as a function of temperature at the (440) structural Bragg peak. Upon cooling through the HO transition, there is no evidence of peak splitting from an orthorhombic distortion. Inset shows the data at 16K and 18K. Diffraction data were taken with $E$ = 8.6 $keV$ }
%\label{Figure:Structure}
%\end{figure}

In this study, we perform high resolution inelastic x ray scattering to study the acoustic phonon modes. Most previous scattering studies focused on magnetic excitations\cite{Broholm:1991vo,Wiebe:2004gz, Wiebe:2007dq, Janik:2009vu, Williams:2012dd, Bourdarot:2014fga}. Neutron scattering data on phonon excitations are complicated due to difficulties separating magnetic from vibrational excitations in most of the data.\cite{Butch:2015ev,Buhot:2015ip}. Because the magnetic scattering cross section for x rays is negligible, x rays are the ideal probe to study phonon excitations.

Inelastic scattering measurements were taken on the HERIX spectrometer at Sector 30 of the Advanced Photon Source at Argonne National Laboratory. The experiment was performed in transmission geometry with a fixed final energy of 23.724 keV. The data were collected during two experiments on the same single crystal sample. In the first of these experiments, scattering from Kapton tape used for sample containment contributed to quasi-elastic and background scattering. In the later experiment, a redesigned sample holder precluded Kapton scattering from satisfying the geometrical requirements to reach the detectors. 

The energy resolution of $\sim1.5$ meV was determined by the FWHM of a pseudo-Voigt quasi-elastic scattering around $\hbar \omega = 0$ on plexiglass. The momentum resolution of $\sim0.012 {\AA}^{-1}$ was determined by measuring the FWHM of the Bragg peaks. Unlike neutron scattering, to good approximation the different components of the resolution function are uncoupled: there is no tilting of the resolution ellipsoid since the change in energy is so small compared the total energy of the x ray. Background counts are primarily due to detector dark counts, which contribute negligibly to the experiment.

The scattering cross section for phonon excitations is proportional to $(\mathbf{Q \cdot \xi})^2$ where $\mathbf{Q}$ is the total momentum transfer and $\mathbf{\xi}$ is the eigenvector of a phonon mode. Thus, by taking measurements at $(400)$ $(220)$ and $(008)$ zone centers, it is possible to independently measure longitudinal and transverse modes along high symmetry directions. Figure~\ref{Figure:Scans} shows scans along $(4-\epsilon, 0, 0)$ and $(2+\epsilon,2+\epsilon,0)$, which correspond to longitudinally polarized modes along high-symmetry directions $\Gamma$ to $\Sigma$ and $\Gamma$ to X.

The intensity for phonon scattering is proportional to the dynamic structure factor $S(\mathbf{Q},\omega)$, which is fit to damped harmonic oscillator (DHO) response function. The scattering from the Kapton tape, which is present in some scans, is accounted for phenomenologically by two additional fit parameters: an amplitude for an overdamped harmonic oscillator and a constant background.

\begin{figure}[ht]
\includegraphics[width=9cm]{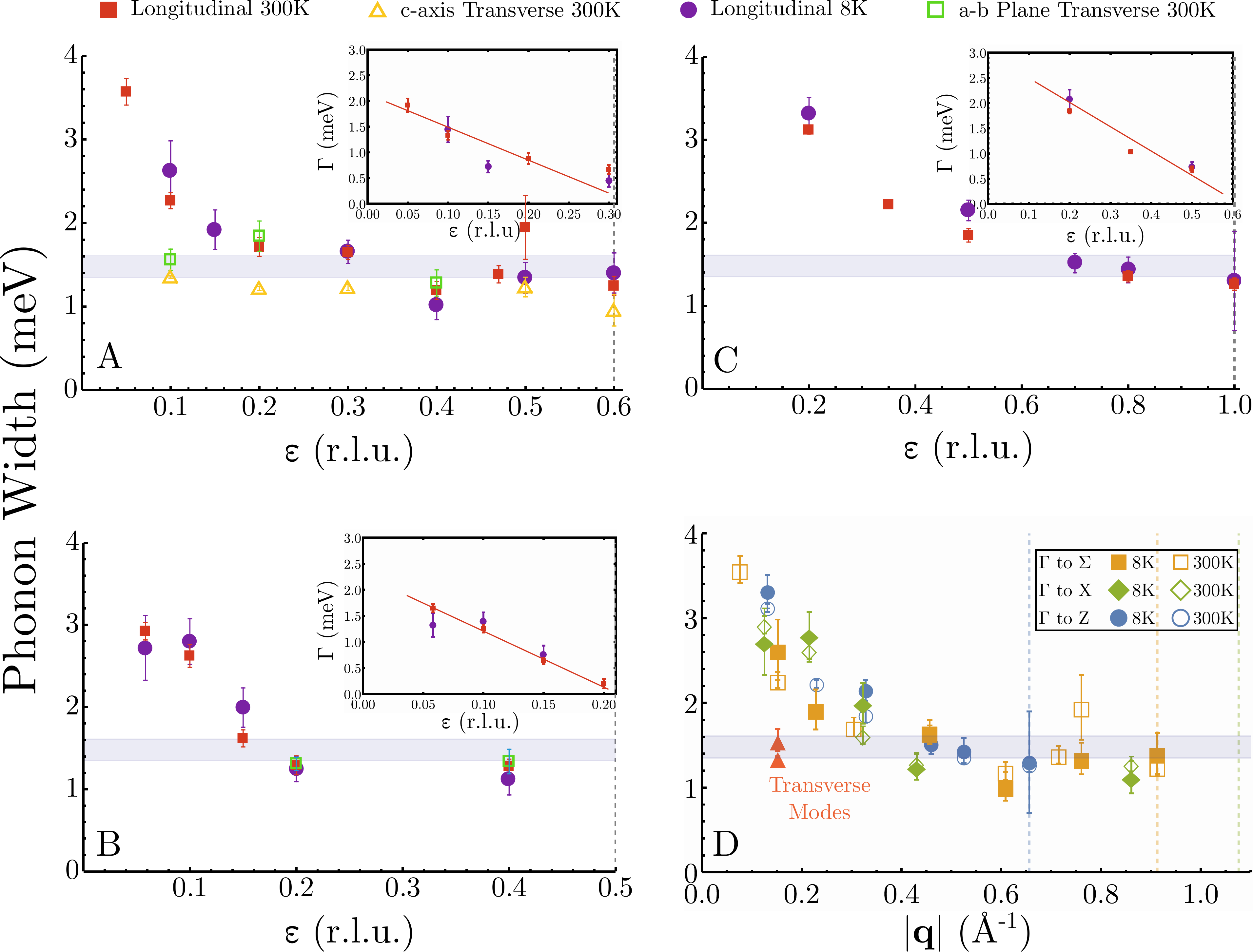}
\caption{Acoustic phonon mode widths from fits to a damped harmonic oscillator. (A) (B) and (C) show cuts along $\Gamma$ to $\Sigma$, $\Gamma$ to X, and $\Gamma$ to Z respectively. The shaded region denotes the FWHM of the energy resolution. Insets show the intrinsic width of longitudinal modes near the zone center. The intrinsic widths are extracted from fits of a damped harmonic oscillator convolved with the resolution function. (D) shows the width of the DHO fits of longitudinal acoustic modes along all three high-symmetry directions as a function of $|\mathbf{q}|$, the reduced momentum transfer of the mode. Transverse modes at (4, 0.1, 0) and (8, 0.1, 0) are included for comparison. Dashed vertical lines denote the zone boundaries.}
\label{Figure:Widths}
\end{figure}

\begin{figure*}[t]
\includegraphics[width=18cm]{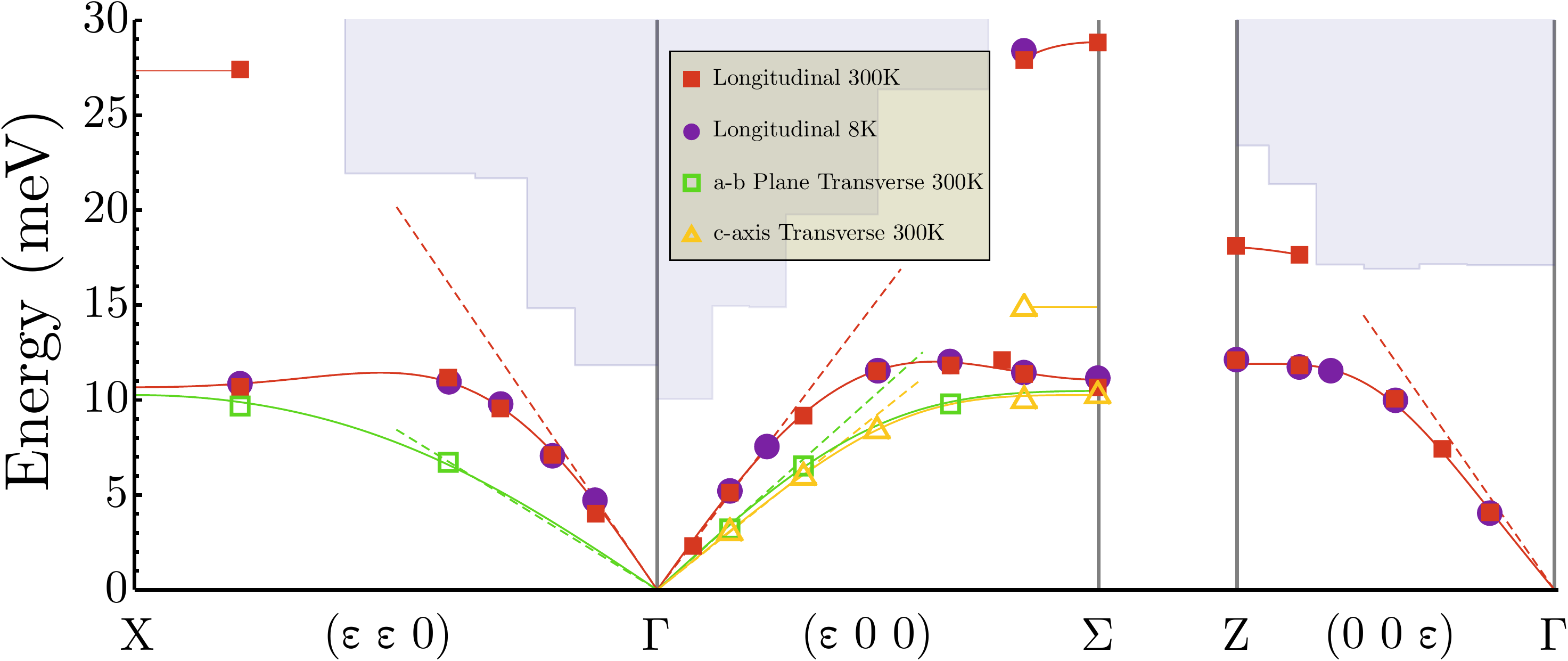}
\caption{Phonon dispersions are shown along various high-symmetry cuts. Solid lines are guide to the eye. Dashed lines are speed of sound calculated from ultrasonic measurements\cite{Yanagisawa:2012fb,Wolf:1994bb}. The shaded region indicates the unexplored region of phase space for longitudinal modes. The Brillouin zone and high-symmetry points are shown in Fig.~\ref{Figure:Scans}(A)}
\label{Figure:Dispersion}
\end{figure*}

The longitudinal acoustic modes show a significant broadening near the zone center.  This effect is seen at both 300K and 8K which suggests the broadening is not from phonon-phonon interactions, which is expected to increase with increasing temperature. Fig~\ref{Figure:Widths} shows the width from fitting the phonons to a DHO The insets show the intrinsic width of the phonon extracted from convolving the DHO with the instrumental resolution function. These nontrivial widths are caused by a reduction of the phonon lifetime as $\mathbf{q}$ approaches zero and is indicative of phonons coupling to other degrees of freedom. In contrast, the transverse modes have long lifetimes and remain resolution limited.

Figure~\ref{Figure:Dispersion} shows the measured dispersions along high-high symmetry directions. The error bars are much smaller than the point size. Included are several optic modes measured near the zone boundary. The light blue shading denotes the region of phase space not covered by the longitudinal scans. Any longitudinal optic modes in the white region can only exist if they have extremely small scattering amplitude. As the atomic form factors for x-ray scattering are proportional to Z, this can only occur for optic modes with eigenvectors dominated by the silicon atoms. These observations are consistent with previous reports.\cite{Butch:2015ev} The dashed lines are the speeds of sound calculated from ultrasound measurements from Yanigisawa et al. along $\Gamma$ to $\Sigma$ and $\Gamma$ to X and from Wolf et al. along $\Gamma$ to Z \cite{Yanagisawa:2012fb,Wolf:1994bb}. We did not measure any temperature dependence to the dispersion between 300K and 8K.

\begin{figure}
\includegraphics[width=8.5cm]{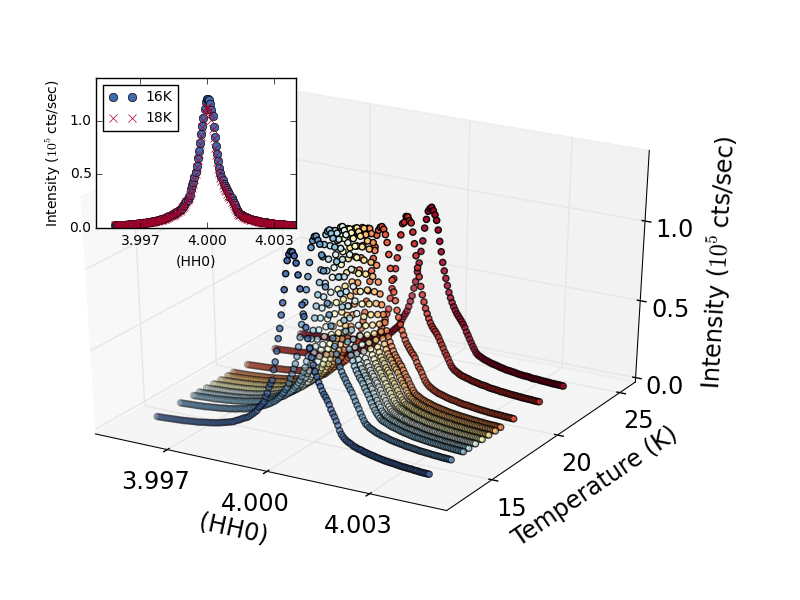}
\caption{ $\theta$-$2\theta$ diffraction data as a function of temperature at the (440) structural Bragg peak. Upon cooling through the HO transition, there is no evidence of peak splitting from an orthorhombic distortion. Inset shows the data at 16K and 18K. Diffraction data were taken with $E$ = 8.6 $keV$ }
\label{Figure:Structure}
\end{figure}

Diffraction on the same sample was performed on the X21 beamline at the National Synchrotron Light Source at Brookhaven National Laboratory. The data show sharp Bragg peaks, which attests to the crystalline quality. Diffraction at the (440) Bragg peak is shown in Fig.~\ref{Figure:Structure}. If an orthorhombic distortion exists, as see in by Tonegawa et al. it would result a splitting of this peak\cite{Tonegawa:1jm}. The lack of observed splitting within our resolution allows us to rule out any orthorhombic distortion greater than $\delta = \frac{a - b}{a + b} = 6*10^{-5}$. This bound is smaller than the distortion reported by Tonegawa et al. in a sample with extremely RRR $\sim$ 670. Our sample has an RRR of $\sim17$ and so the lack of a distortion agrees with the observed RRR dependence reported by Tonegawa et al and is consistent with diffraction data by Tabata et al.\cite{Tabata:2014is}

Both Sharma et al. and Behnia et al. see a large increase in the thermal conductivity, both in-plane and along the $c$ axis, upon cooling through the HO transition. By analyzing the thermal hall conductivity, they conclude that the electronic contribution is extremely small and therefore the lattice contribution must be large and change significantly upon entering the HO phase. \cite{Behnia:2005iy,Sharma:2006fa}

Under a relaxation time approximation, the thermal conductivity from an excitation of wave vector $\mathbf{q}$ and branch $j$ is given by
\begin{align}
\label{Equation:Conductivity}
\kappa_{\mathbf{q},j}=\frac{1}{3}C_{\mathbf{q},j} \mathbf{v}_{\mathbf{q},j}\lambda_{\mathbf{q},j}
=
\frac{1}{3}C_{\mathbf{q},j} \mathbf{v}_{\mathbf{q},j}^2 \tau_{\mathbf{q},j}
\end{align}
where $C_{\mathbf{q},j}$ is the heat capacity, $\mathbf{v}_{\mathbf{q},j}$ is the group velocity, $\lambda_{\mathbf{q},j}$ is the mean free path, and  $\tau_{\mathbf{q},j}$ is the relaxation time. We can extract the relaxation time from the width of the phonons, $\Gamma$ as $\tau = \frac{h}{\Gamma}.$\cite{Maradudin:1962ub} This method has successfully been used to calculate thermal conductivity from scattering data \cite{Pang:2013dn}.
%insert reference to Phys. Rev. 128, 2589 ? Published 15 December 1962

From inspection of Eq.~\ref{Equation:Conductivity}, at temperatures near $T_0$, only acoustic phonons near the zone center have both sufficiently high heat capacity and group velocity to contribute significantly to the thermal transport. The acoustic modes away from the zone center have a very small group velocity and so contribute little; the optic modes are too high of energy to be thermally populated and so therefore should also contribute negligibly.

The measured intrinsic line width of the longitudinal branches allow for the direct approximation of their contribution to the thermal transport. For ease of calculation, we approximate the dispersion as isotropic. Deviations from this approximation do not affect the calculation much as mode contribution falls off as $\cos(\theta)^2$, where $\theta$ is the angle between the mode propagation and thermal transport. The Brillouin zone is approximated as a Debye sphere.

For the region of the dispersion near the zone center, for which the intrinsic line width can be extracted, we use this to calculate the mean free path. For the remainder of the branch, we use the upper bound of 1mm, which represents boundary scattering from sample sizes typical for thermal transport measurements. This then represents an upper limit for what the branch can contribute. For thermal transport along the $a$ axis, the longitudinal mode only contributes $0.2 \frac{\mathrm{W}}{\mathrm{K \; m}}$, a negligible amount of the total conductivity. The mean free path of a phonon is $\lambda = (\partial \hbar \omega / \partial q) (1/\Gamma)$. Because only modes with intrinsic widths on the order of 1meV are measurable beyond resolution, and since the speeds of sound are $\sim$30 meV \r A, any modes with with measured width outside of resolution have a mean free path only on the order of tens of angstroms. Hence, it is reasonable that the longitudinal branch contributes such a small amount.

For the transverse modes, we cannot extract an intrinsic line width from any of the modes as they are resolution limited. Therefore, it is impossible to rule out a significant increase in the thermal transport from these modes upon entering the Hidden Order. However, the mean free path lower bound for transverse phonons near the zone center is only $\sim$30 \r A as anything smaller would induce a measurable width. %Since the minimum intrinsic width for which the phonon will be non-resolution limited is $\sim$1meV, the mean free path of transverse modes near the zone center has a lower bound $\sim$ 50 \r A.  
We can calculate an effective mean free path for the entire transverse branches necessary to match the reported thermal conductivity. This effective $\lambda$ must increase from $\sim$ 1 $\mu$m above the HO to $\sim$ 50 $\mu$m immediately below it. This change must occur to the mean free path throughout the entire Brillouin zone to account for the increase of thermal conductivity in all measured directions and thus seems an unlikely source for the increase in thermal conductivity.

A more plausible explanation is that the known magnetic excitation are the source of the thermal transport increase in the HO phase. Thermal transport from magnetic excitations has been observed in a variety of materials including multiferroics, spin ice, monolayer cuprates, and low-dimensional spin systems.\cite{Matsuoka:2014er,Berggold:2006io,Zhao:2011dd,Toews:2013ij,Hofmann:2003gs}   Neutron scattering data reveal that below $T_0$, there are well-defined magnetic excitations throughout the Brillouin zone. Lowest energy excitations occur at ${\mathbf{Q_0} = (1~0~0)}$ and ${\mathbf{Q_1} = (1.4~0~0)}$ with gaps of $1.7$ and $4.2$~meV respectively. \cite{Bourdarot:2014fga} These modes are dispersive and sharp in all measured directions and are low enough in energy to be thermally active and thus must contribute to thermal transport.

Above $T_0$,the excitation at $\mathbf{Q_0}$ is thought to become gapless but heavily overdamped. The excitation at $\mathbf{Q_1}$ lowers to $2.1$~meV, but also becomes heavily damped \cite{Bourdarot:2014fga}. Thus the magnetic excitations above $T_0$ are incapable of carrying appreciable thermal current.

The broadening of longitudinal acoustic phonon as $\mathbf{q} \rightarrow 0$ suggest the modes are coupling to other degrees of freedom at all temperatures. One possibility is coupling to magnetic excitations at $Q_0$, which becomes a zone center for a simple tetragonal unit cell proposed as the true unit cell.\cite{Meng:2013hj,Hassinger:2010wk} Though the magnetic excitations change considerably upon entering the HO, magnetic scattering exists at $\mathbf{Q_0}$ above the transition and so may still couple to phonons. Electronic Raman spectroscopy has recently revealed a $\mathbf{q}=0$ excitation which may also be a source of phonon coupling.\cite{Kung:2015ip} Finally, it is possible that part of the broadening comes from disorder. Regardless of the exact nature of the coupling, it is clear that low energy electronic and/or magnetic excitations play a significant role in the low temperature thermal properties and have a strong coupling to the phonons at small wave vectors.

The work at MIT was supported by the Department of Energy (DOE-BES) under Grant No. DE-FG02-07ER46134. Use of the Advanced Photon Source at Argonne National Laboratory was supported by the DOE-BES under Contract No. DE- AC02-06CH11357. The construction of HERIX was partially supported by the NSF under Grant No. DMR- 0115852.

%The existence of magnetic excitations carry thermal current has been seen in low-dimensional materials. This is, to our knowledge, the first example in which magnetic excitations have been demonstrated to carry thermal current in 3-D materials.

\bibliography{bibliography}
\bibliographystyle{phjcp}

\end{document}